# 3D digital reassembling of archaeological ceramic pottery fragments based on their thickness profile

**Michail I.Stamatopoulos[††] and Christos-Nikolaos Anagnostopoulos[†]**

[†]Social Sciences School, Cultural Technology and Communication Dpt. University of the Aegean, Greece, Email: canag@aegean.gr
[††]School of Science and Technology, Computer Science Dpt. Hellenic Open University, Greece, Email: std069191@ac.eap.gr

**Abstract:** The reassembly of a broken archaeological ceramic pottery is an open and complex problem, which remains a scientific process of extreme interest for the archaeological community. Usually, the solutions suggested by various research groups and universities depend on various aspects such as the matching process of the broken surfaces, the outline of sherds or their colors and geometric characteristics, their axis of symmetry, the corners of their contour, the theme portrayed on the surface, the concentric circular rills that are left during the base construction in the inner pottery side by the fingers of the potter artist etc. In this work the reassembly process is based on a different and more secure idea, since it is based on the thickness profile, which is appropriately identified in every fragment. Specifically, our approach is based on information encapsulated in the inner part of the sherd (i.e. thickness), which is not -or at least not heavily- affected by the presence of harsh environmental conditions, but is safely kept within the sherd itself. Our method is verified in various use case experiments, using cutting edge technologies such as 3D representations and precise measurements on surfaces from the acquired 3D models.

## 1. Introduction

In every archaeological excavation, a variety of small ceramic pottery fragments (or called sherds) are revealed, which will provide valuable data for the excavation itself. The nature of these small objects give them properties which make them highly resistant to time and wear. Moreover, sherds and these small fragments cannot be do not stolen, but and they remain near to the place where the ceramic vase is destroyed. Therefore the information carried within remain undeletable over the centuries, yet difficult to reassembled. The amount and the high value of information that is encapsulated within, rightly gives to the sherds, the title of best data carrier from ancient times to our days.
For the reassembly of a ceramic vase vessel, the majority of the scientific methods suggested by various research groups and universities, depend on the outline of sherds, or matching the broken surfaces, the colours, geometric characteristics, the axis of symmetry, the corners of the contour, or even on the theme portrayed on it [1-8]. All these techniques suffer from the problems introduced by the external wear and decay of the material during the exposure in the soil. As a result, many characteristics on the surface of the sherd will be altered during the long time, leading to severe decrease in the efficiency of research methods based on external characteristics [8-10]. Therefore, it is no surprise that, to our knowledge and after extensive search in museum laboratories in countries with high archaeological interest (Greece, Italy, Vatican, Egypt), an automatic method that works effectively to restore a true ancient theme is not available. In contrast, the process is performed manually by experts in the field, aligning the fragments by minimizing the distance between their adjoining regions while simultaneously trying to ensure geometric continuity across them (if possible).

## 2. The Thickness Profile (TP) method

The proposed methodology is based on a new approach that can be considered scientifically more effective than others, as we seek important information encapsulated inside the core of the sherd and not in its surface. Our approach works even if some of the sherds are missing, it is not affected by the presence of external wear and damages, nor by the geometrical shapes or by the colour degradation of the pottery.

The new method is based on exploration, extraction and utilization of all possible thickness information (i.e. Thickness Profile) which may have encapsulated inside each sherd. All this information, as a sequence of numbers, can be sorted, compared and provide a complete and efficient solution for the complex problem of reconstruction, reassembling, restoration and recovery of an ancient broken ceramic pottery which is fragmented into pieces, of random sizes and shapes.

The basic idea is based on the fact that as the artist rotates the pliable clay on the wheel to create the pottery, it creates distinguishing thickness measurements, which are unaltered and can easily be detected. The gradual construction of a pottery on a wheel starts from the base, it continues up to the main body and usually ends at the neck and at the rim. This gradual upward movement of the artist in the clay body, creates a certain thickness profile which can be detected, as it varies, with absolute certainty, from point to point, from edge to edge and of course it differs between different artefacts. This distinction creates the image of a structure which resembles a stack of horizontal rings [11-14], with specific thickness in relation to the pottery height (Fig. 1). Specifically, as the fingers of artist push outwards (expansion of the clay body), the clay in these points is getting thinner, while in the case of inward pressure (contraction of the clay body) the clay is getting thicker.

Based on the above, it is reasonable to consider, that each sherd can theoretically fit to a specific point of the stack of rings and hence to a corresponding point in the overall thickness profile or thickness contour of a particular pottery.

The proposed methodology comprises of three basic steps, namely: i) the 3D scanning of each fragment, ii) the extraction of their optimal thickness profile (TP) and iii) the repetitive process for maximizing matching scores between TPs in order to achieve a locally optimal alignment between possibly neighbouring sherds.

### 2.1 Creation of 3D models (points clouds)

As it is absolutely necessary, to obtain the appropriate profile measurements without damaging the sherds, our method is implemented and validated using their identical 3D models, making the extraction of measurements fully reproducible and non-intrusive to the sherd itself (Fig. 2).

Each sherd is placed on a stable basis and photographed panoramic, from close distance, from all sides and from various angles. The result is a set of 30 photos for each sherd, which is transformed into a 3D model (point clouds and mesh), using a specialized software (Fig. 3).

### 2.2 Thickness profile (TP) extraction

It is very important for each sherd, to detect the plane with the richest information concerning thickness. The highest vertical plane to the horizontal rings is selected that allows the extraction of the maximum possible thickness profile. This is especially important, since the best possible thickness profile for each sherd is needed, in order to perform the optimum thickness matching score between neighbouring sherds. Fig. 4 shows an example of finding the appropriate plane.

This vertical flat plane is absolutely oriented with the horizontal inner lines of the sherd and using an appropriate 3D modelling software TP of the sherd is calculated accurately. For the calculation, we perform thickness sampling for every 1 mm (Fig. 5).

### 2.3 Thickness Profile matching

Following the previous steps, by the process of sliding small thickness profiles across larger ones until the achievement of optimum fit, the method retrieves candidate matches between sherds performing local score optimization. Examples of Thickness Profiles are shown in Figure 6.

The Thickness Profile method, is a semi-automatic method, as in the case of two adjoining sherds, it cannot decide which one will be placed left and which will be placed right. At this point the expert eye of the archaeologist-user, should give the correct arrangement (left or right). In addition, small sherds usually do not have the ability to give adequate thickness information, and therefore the probability for erroneous arrangement is increased. For these reasons, the process is done in stages, starting from the largest available sherd and moving gradually to the smaller ones.

Starting from a “master” sherd, the main target is to stack progressively more and more thickness information and thus greater thickness profile, increasing the chances to match the remaining sherds in the right place.

Hence, at the beginning of the process, the largest available sherd is assigned with the role of the “driver” (master sherd) and as the process is executed and two or more pieces are matched together, a single meta-sherd is formed with an increased Thickness Profile (Fig.7). Intuitively, this corresponds to virtually gluing sherds together, while the matched parts of TPs become common values to TP of the meta-sherd. This approach ensures that parts of TPs that have previously been matched in the some pair of sherds are no longer considered for future pairings.

The matching procedure looks like a “fluctuation” smaller thickness profiles on larger ones. The search for the ideal match, could be found only if two profiles fit perfectly together and the sequential numbers of the small sherd match somewhere in the sequence of numbers of the large sherd. The perfect match would be possible only in an ideal case and not on real data. Apart from the above, our method is based on the plurality of measurements and less in the accuracy of measurements. Therefore, the method is searching for the highest “score”, with the fewer differences in most possible comparisons. The best “score” is defined as the sum of the absolute differences between the more possible comparisons between two sherd profiles.
The Thickness Profile method works better for pottery vessels, that were built on recently archaeological periods (Archaic, Classical, Hellenistic period) as these vessels were manufactured with finest “technology” characteristics (better wheels, clay quality, best cook, etc.). As a result they feature more sophisticated geometric features and are characterized by distinctive thickness information that can be exploited by the proposed method.

## 3. Experimental validation

Using exact hand-made replicas of ancient ceramic potteries as the driving example, we demonstrate in this section the accuracy and efficiency of the proposed method. The vessels are intentionally broken and part of the resulting material are digitally reassembling successfully by the Thickness Profile method.
Before breaking them, horizontal and vertical lines are marked in the surface, which will be later used and confirm our methodology. Specifically, every vessel was placed on a horizontal surface and concentric circular contours (every 0.5 cm) were drawn in its external surface. Furthermore the outer surface is partitioned into eight different areas (i.e. A, B, C, D, E, F, G and H) using vertical lines at 45, 90, 135, 180, 225, 270, 315 and 360 degrees. The latter marking procedure was confirmed by lighting a vertical laser beamer on and along the outer surface of the vessel at the above mentioned angles. An example of an experimental object before breaking is shown in Figure 8.
For the shake of simplicity, we demonstrate the efficiency of TP method using five neighboring sherds (A4, A5, B10, C2, C15) of and intentionally broken replica of an ancient pottery. Figure 9 demonstrates the acquired 3D models of those fragments and the extraction of the optimal plane for the calculation of the distinctive thickness measurements. Then, the 3D software performs thickness sampling for every millimeter and the thickness profile for each plane is created. Table I indicates the number of acquired thickness measurements for each sherd in this example, while their Thickness Profiles (TP) are depicted in Table II.

**Table I.** Sherd id and number of thickness measurements.

| Sherd id | # of acquired thickness measurements (per mm) / color convention in Figures 10 and 11 |
|---|---|
| A4 | 61 / fuchsia |
| A5 (master) | 57 / red |
| B10 | 36 / blue |
| C2 | 17 / purple |
| C15 | 15 / green |

**Table II.** The Thickness Profiles (in mm) of the five sherds as measured in the respective 3D models.

| Measurement id | C15 TP (mm) | B10 TP (mm) | A5 TP (mm) | A4 TP (mm) | C2 TP (mm) |
|---|---|---|---|---|---|
| 1 | 5,26 | 5,81 | 5,94 | 5,44 | 4,94 |
| 2 | 5,2 | 5,74 | 5,94 | 5,56 | 4,98 |
| 3 | 5,22 | 5,72 | 5,86 | 5,41 | 4,96 |
| 4 | 5,21 | 5,68 | 5,8 | 5,44 | 5,02 |
| 5 | 5,1 | 5,62 | 5,74 | 5,28 | 5,07 |
| 6 | 4,96 | 5,63 | 5,73 | 5,34 | 5,22 |
| 7 | 4,9 | 5,54 | 5,62 | 5,34 | 5,23 |
| 8 | 5,02 | 5,53 | 5,46 | 5,46 | 5,2 |
| 9 | 5,6 | 5,38 | 5,42 | 5,56 | 5,13 |
| 10 | 5,72 | 5,3 | 5,38 | 5,36 | 5,15 |
| 11 | 5,86 | 5,24 | 5,32 | 5,34 | 5,21 |
| 12 | 5,86 | 5,26 | 5,23 | 6,33 | 5,32 |
| 13 | 5,86 | 5,24 | 5,13 | 5,38 | 5,32 |
| 14 | 5,72 | 5,22 | 5,01 | 5,44 | 5,34 |
| 15 | 5,7 | 5,21 | 4,97 | 5,4 | 5,46 |
| 16 | | 5,15 | 4,83 | 5,4 | 5,39 |
| 17 | | 5,1 | 4,77 | 5,33 | 5,31 |
| 18 | | 5,08 | 4,81 | 5,26 | |
| 19 | | 4,86 | 4,88 | 5,06 | |
| 20 | | 4,81 | 4,95 | 4,99 | |
| 21 | | 4,9 | 5,01 | 5,12 | |
| 22 | | 4,94 | 5,12 | 5,8 | |
| 23 | | 5,02 | 5,24 | 6,07 | |
| 24 | | 5,08 | 5,22 | 5,98 | |
| 25 | | 5,19 | 5,14 | 5,91 | |
| 26 | | 5,28 | 5,1 | 5,88 | |
| 27 | | 5,26 | 5,11 | 5,8 | |
| 28 | | 5,16 | 5,18 | 5,7 | |
| 29 | | 5,16 | 5,32 | 5,56 | |
| 30 | | 5,16 | 5,3 | 5,56 | |
| 31 | | 5,28 | 5,28 | 5,46 | |
| 32 | | 5,3 | 5,3 | 5,37 | |
| 33 | | 5,33 | 5,32 | 5,31 | |
| 34 | | 5,33 | 5,28 | 5,37 | |
| 35 | | 5,36 | 5,34 | 5,5 | |
| 36 | | 5,23 | 5,21 | 5,57 | |
| 37 | | | 5,24 | 5,61 | |
| 38 | | | 5,27 | 5,36 | |
| 39 | | | 5,25 | 5,21 | |
| 40 | | | 5,26 | 5,14 | |
| 41 | | | 5,28 | 5,16 | |
| 42 | | | 5,2 | 5,86 | |
| 43 | | | 5,06 | 6,04 | |
| 44 | | | 5,01 | 6,01 | |
| 45 | | | 5,08 | 6,13 | |
| 46 | | | 5,52 | 6,18 | |
| 47 | | | 5,83 | 6,16 | |
| 48 | | | 6,07 | 6,16 | |
| 49 | | | 5,99 | 6,18 | |
| 50 | | | 5,96 | 6,16 | |
| 51 | | | 5,86 | 6,2 | |
| 52 | | | 5,66 | 6,26 | |
| 53 | | | 5,54 | 6,34 | |
| 54 | | | 5,39 | 6,74 | |
| 55 | | | 5,33 | 6,94 | |
| 56 | | | 5,32 | 7,08 | |
| 57 | | | 5,32 | 7,13 | |
| 58 | | | | 7,23 | |
| 59 | | | | 7,56 | |
| 60 | | | | 7,58 | |
| 61 | | | | 7,62 | |

Using the repetitive procedure described in section 2.3, the results validate that our method allows for accurate reassembly of the five sherds to be achieved with minimal human interaction. The human interaction involves the correct placement of the fragments to the left of to the right of the master sherd. Figures 10 and 11 demonstrate the effective matching of the five sherds in terms of selected planes and thickness profiles respectively.

Finally, Figure 12 presents the final reassembled external surface of the pottery along with a thickness scale representing the common thickness profiles of the assembled fragments. It should be noticed, that the proposed method is effective even if small pieces of the vessel are still missing. This has significant implications for archeology since until now, manual reassembly is usually based on contour-based methods that exploits local surface characteristics in the fragments. However, if such small parts are missing or are altered, severe problems in the reassembly process are imposed.

## 4. Conclusions

In this paper we have presented a new digital reassembly method for ancient ceramic pottery based on 3D models of their fragments and the exploitation of their thickness profile. The results show that our method allows for accurate reassemblies to be achieved with minimal expert interaction. Our approach is based on thickness, which is a kind of information that is encapsulated in the inner part of the sherd that cannot be affected by the presence of harsh environmental conditions. Our method is verified in hand-made ancient ceramic pottery replicas that were broken intentionally into their fragments. Using 3D representations and precise measurements in their surfaces, we demonstrated the validity of our method.

To our knowledge, we have introduced a method that bridges the gap between top-down and bottom-up approaches and answers difficult problems in the excessively time consuming task of manual reassembly of ancient pottery. We intend to expand and fine-tune our methodology with more experiments using specific objects of the archaic period.

# Appendix: List of figures

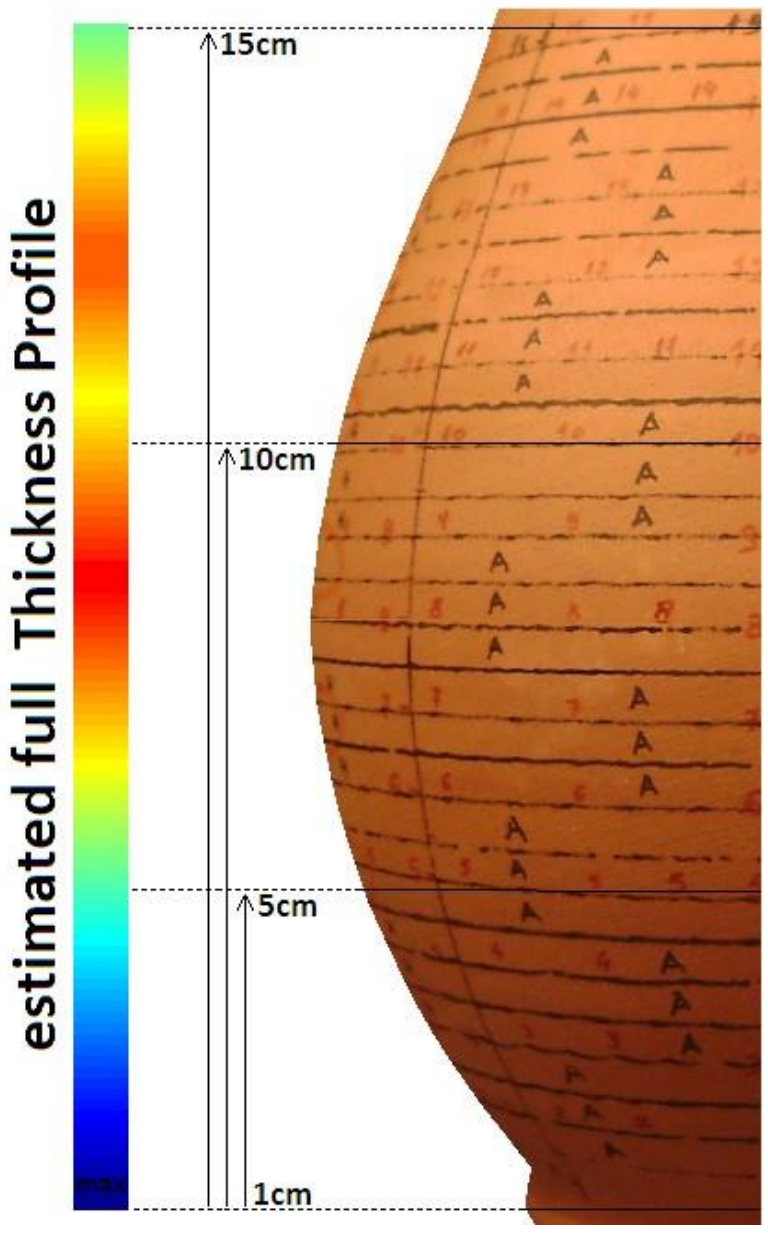

**Fig. 1**. A graphical example of the thickness profile in a pottery vessel. Note the gradually change of the thickness from the base to the main body and up to the neck. Blue and red colours indicate thick and thin profiles respectively.

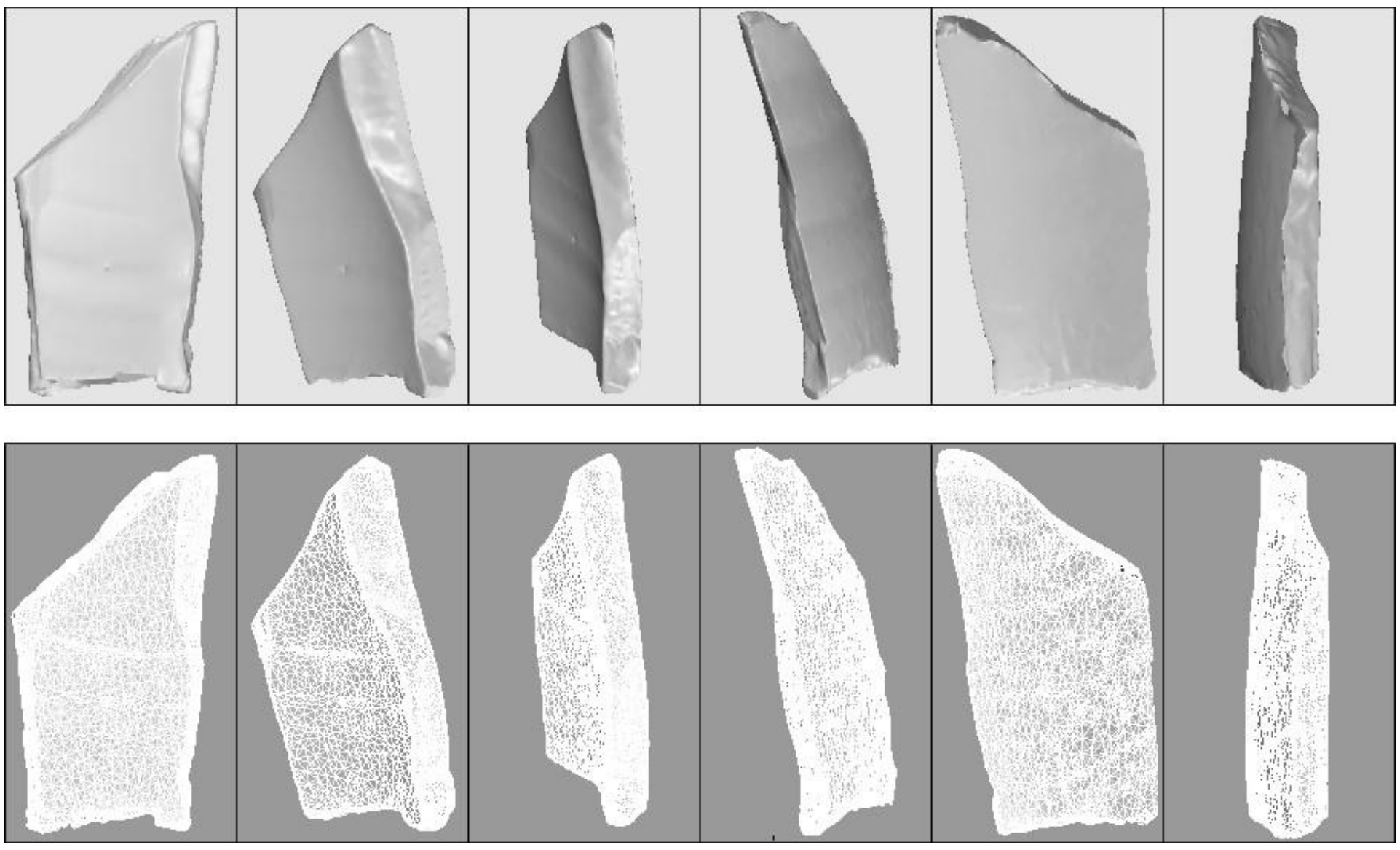

**Fig. 2**. Multiple views of a 3D model of an original sherd.

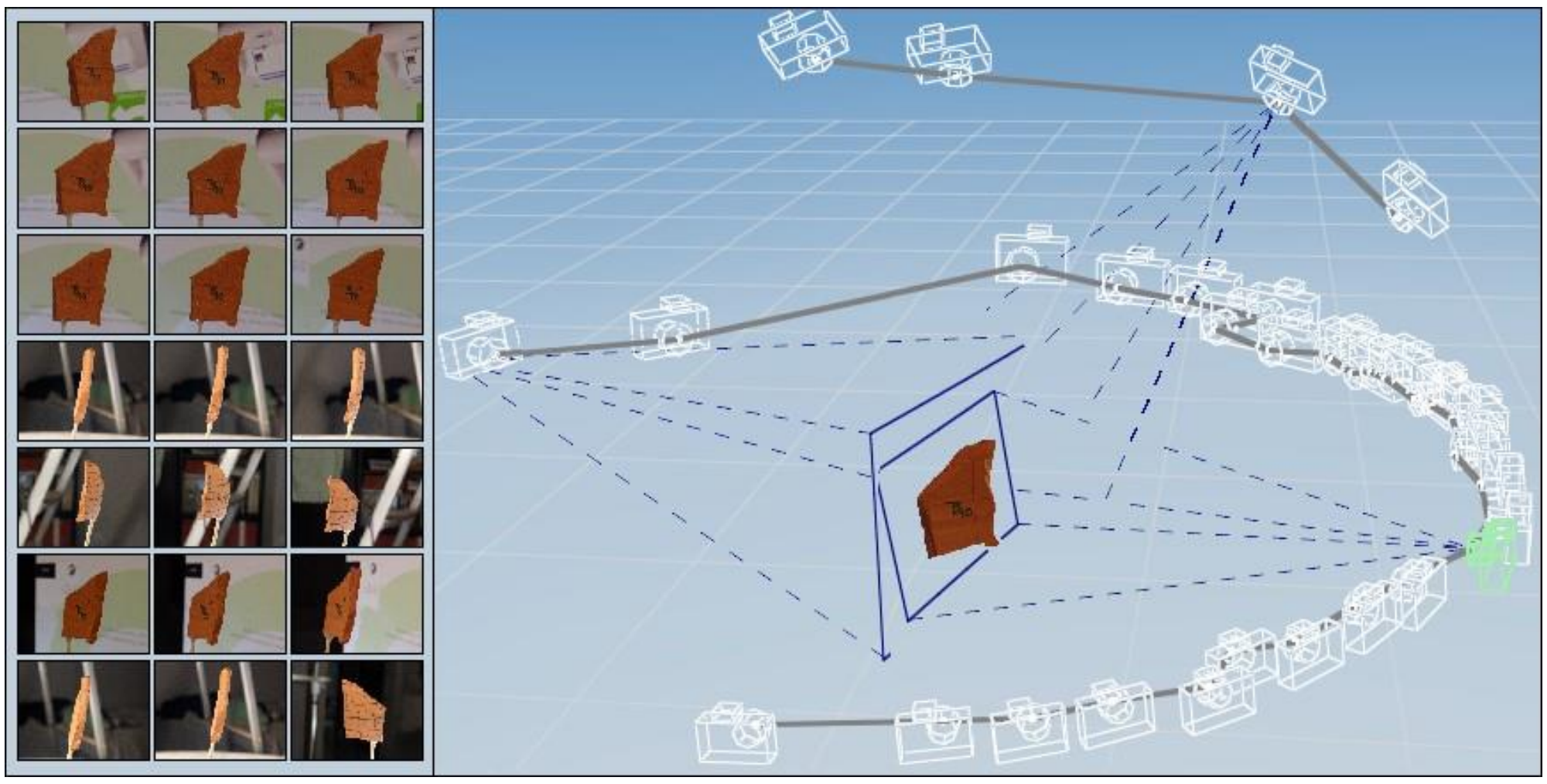

**Fig. 3.** The process of converting the sets from sherds photos into 3D models.

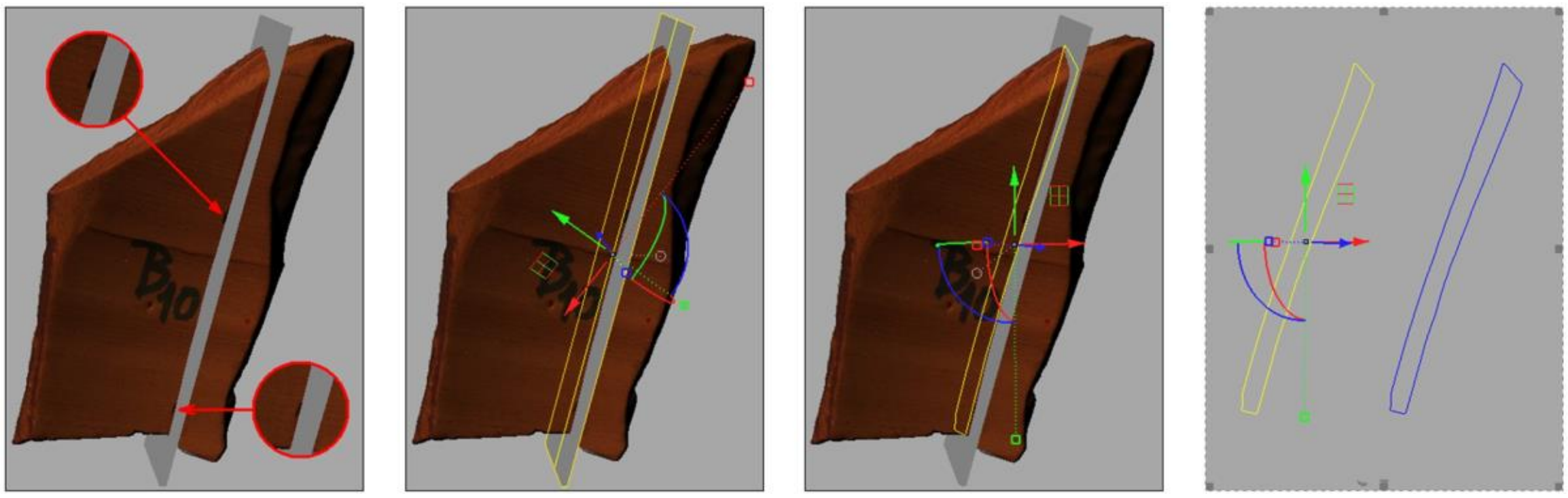

**Fig. 4**. Finding the optimal plane for TP acquisition.

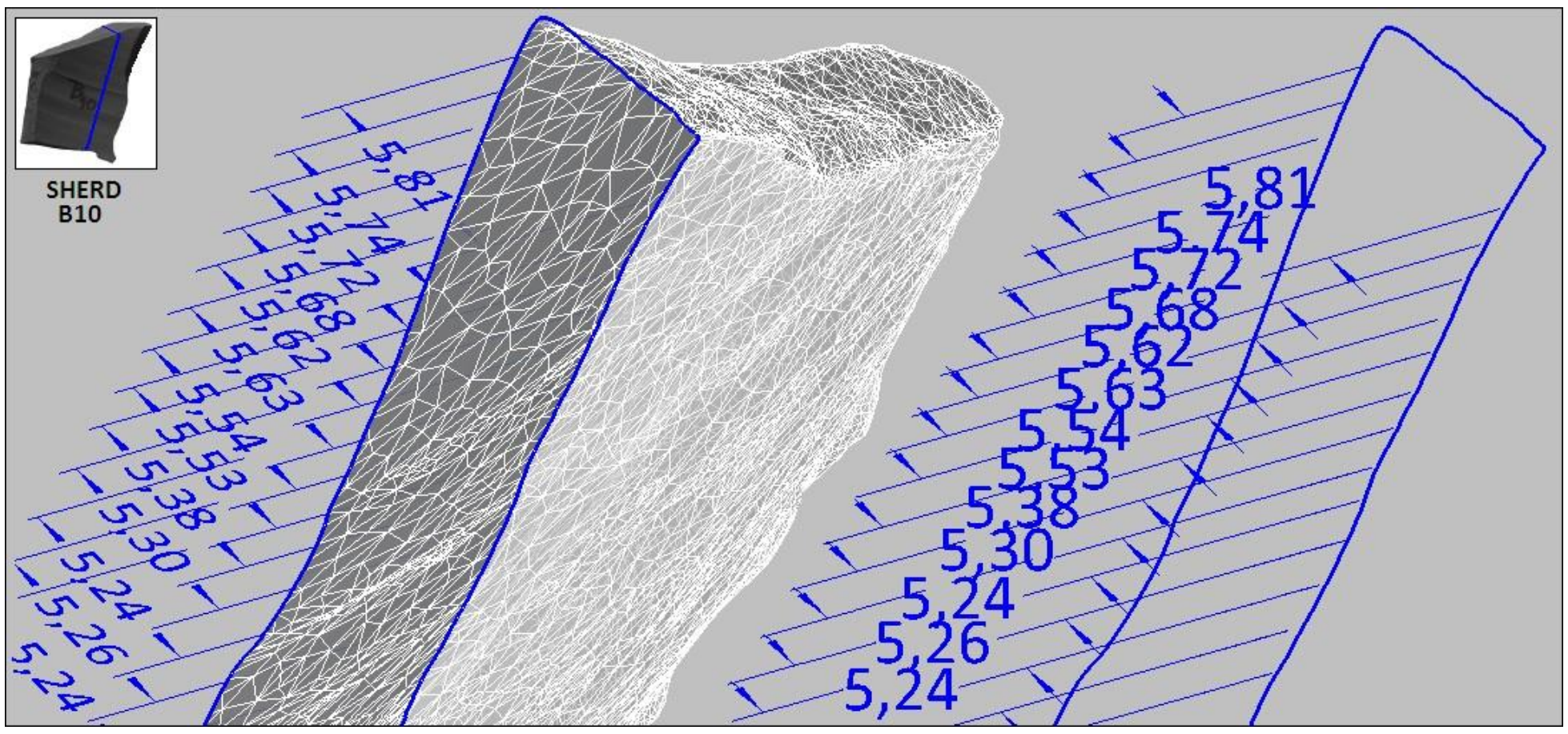

**Fig. 5**. Thickness data acquisition.

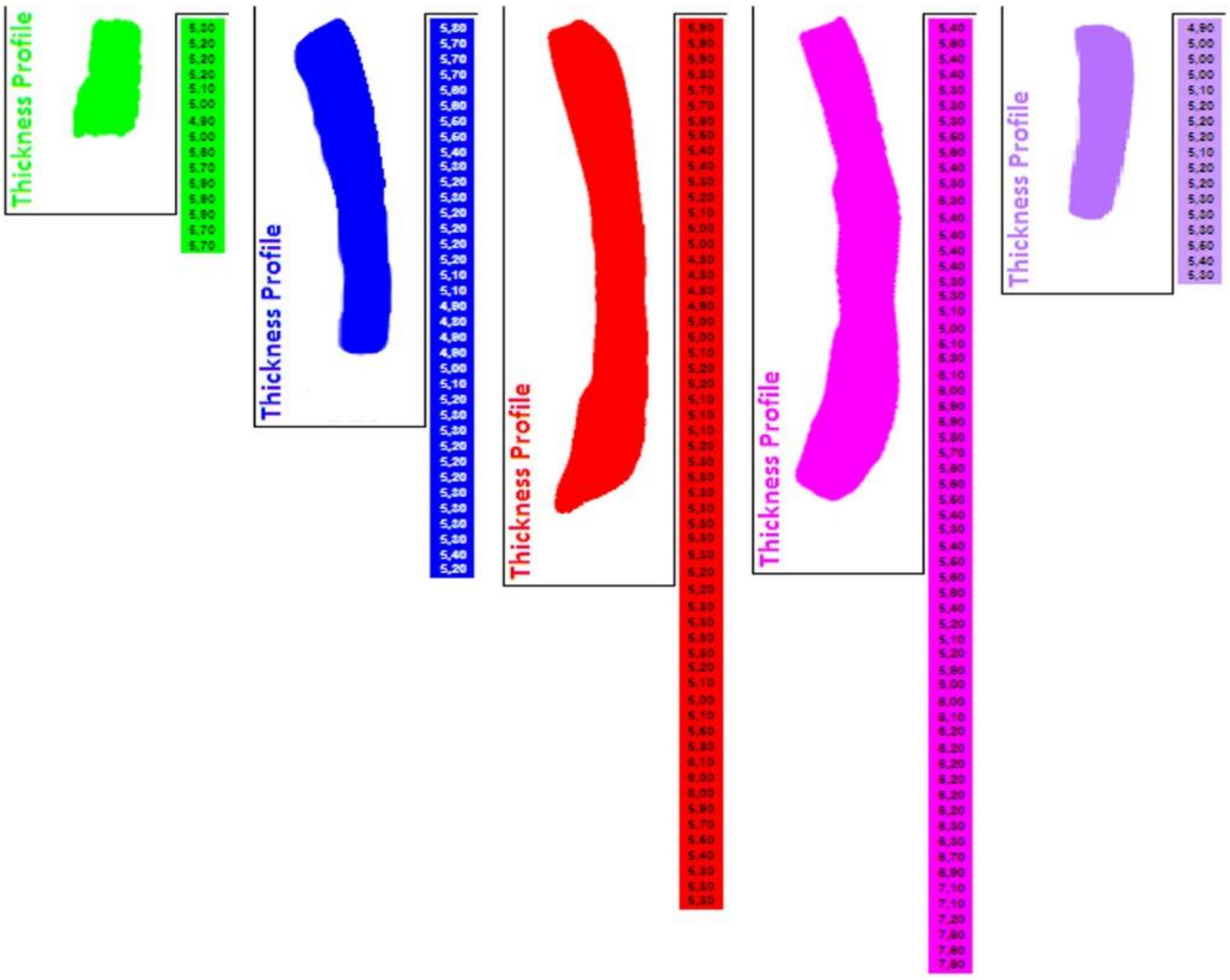

**Fig. 6.** Thickness plane and their respective profiles as a sequence of numbers from several sherds. The profiles will be then compared and candidate neighboring sherds will be matched.

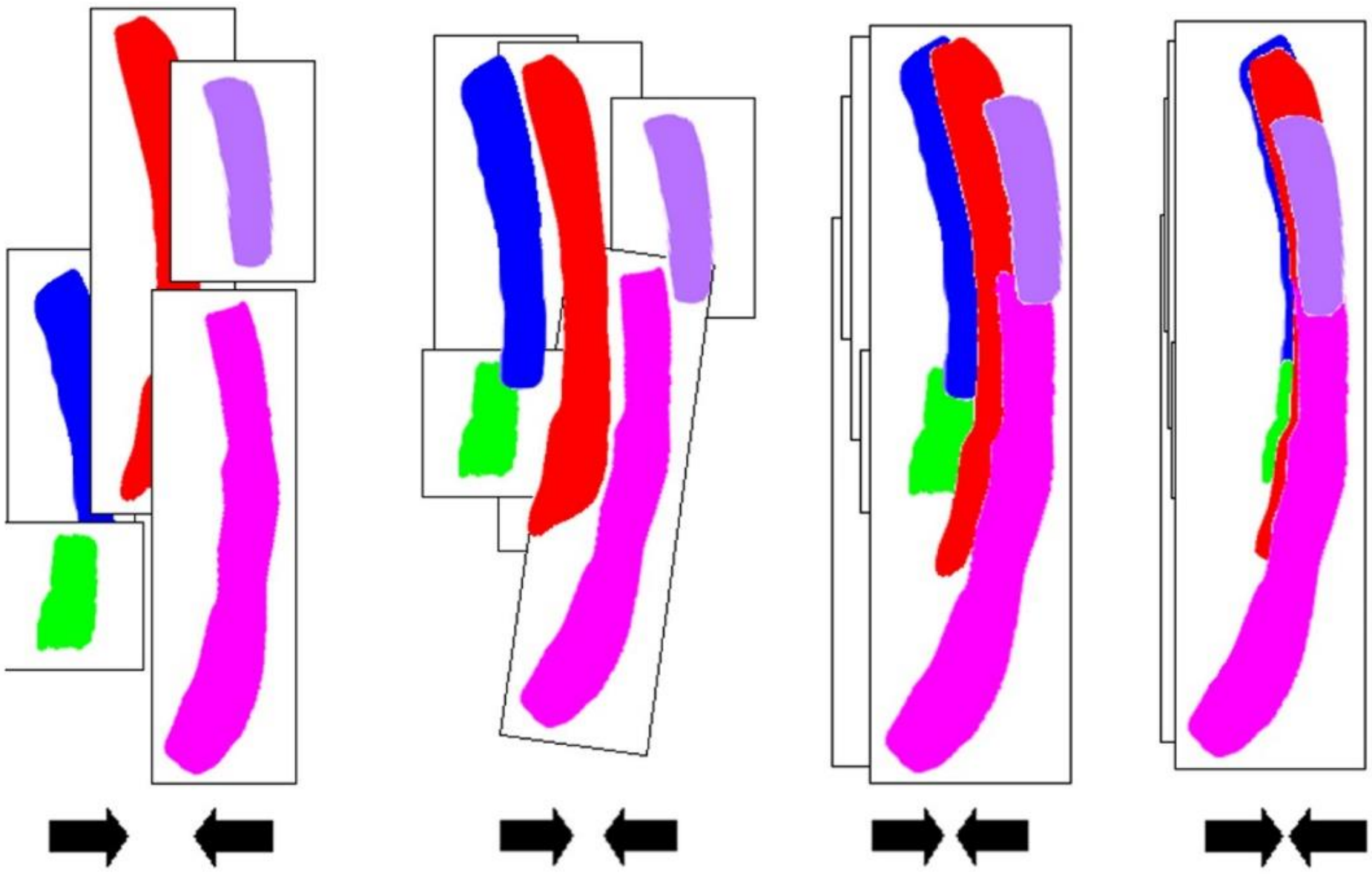

**Fig. 7.** Group of sherds with the thickness profiles side by side. Rightmost as stacked, the expected matching of their outline profiles and the creation of a new meta-sherd.

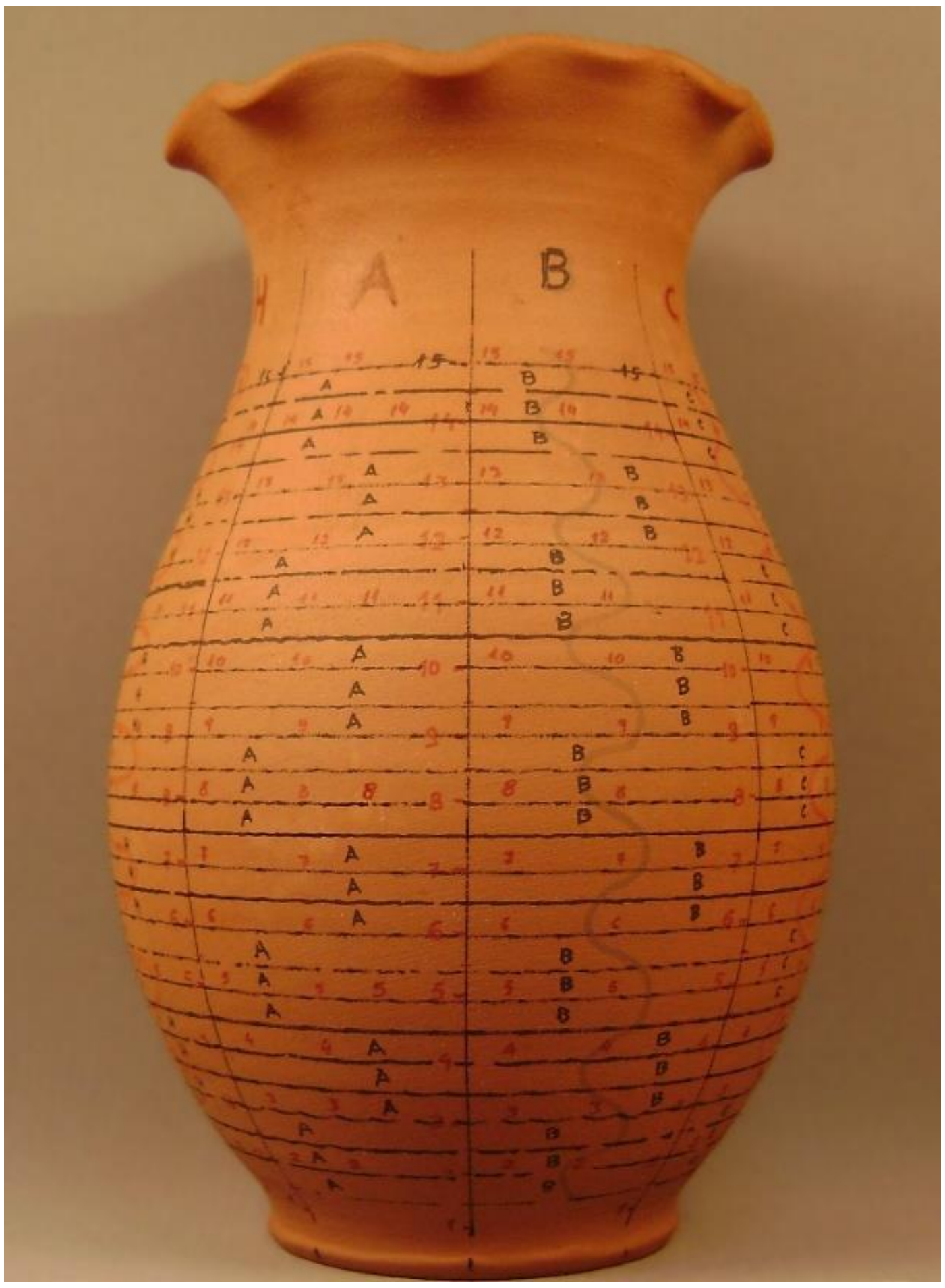

**Fig. 8**. A vessel used for our experiments before breaking.

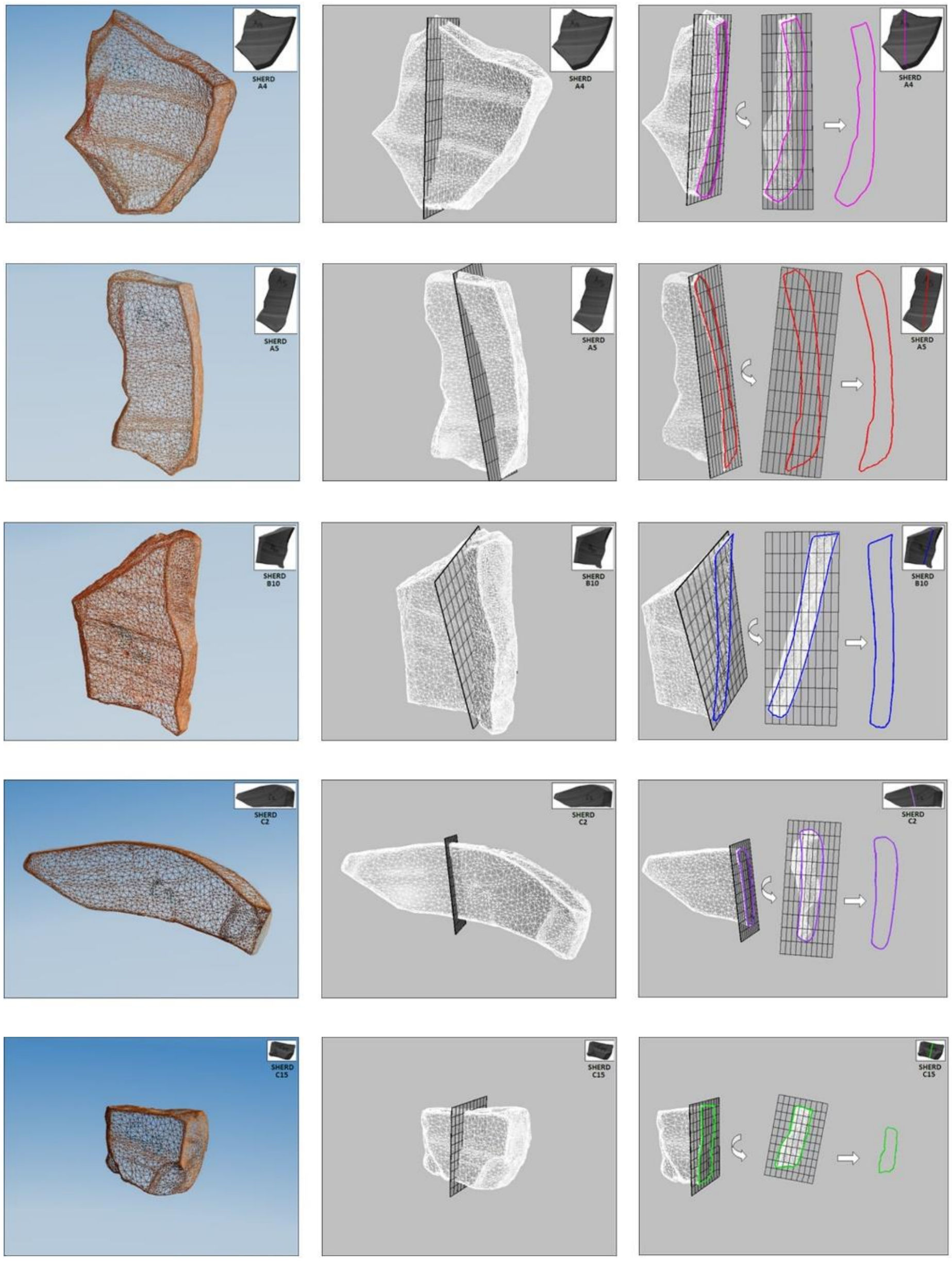

**Fig. 9.** Left column: the 3D models from some sherds, Middle column: the 3D model with the best profile plane position, Right column: the acquisition of the maximum thickness profile.

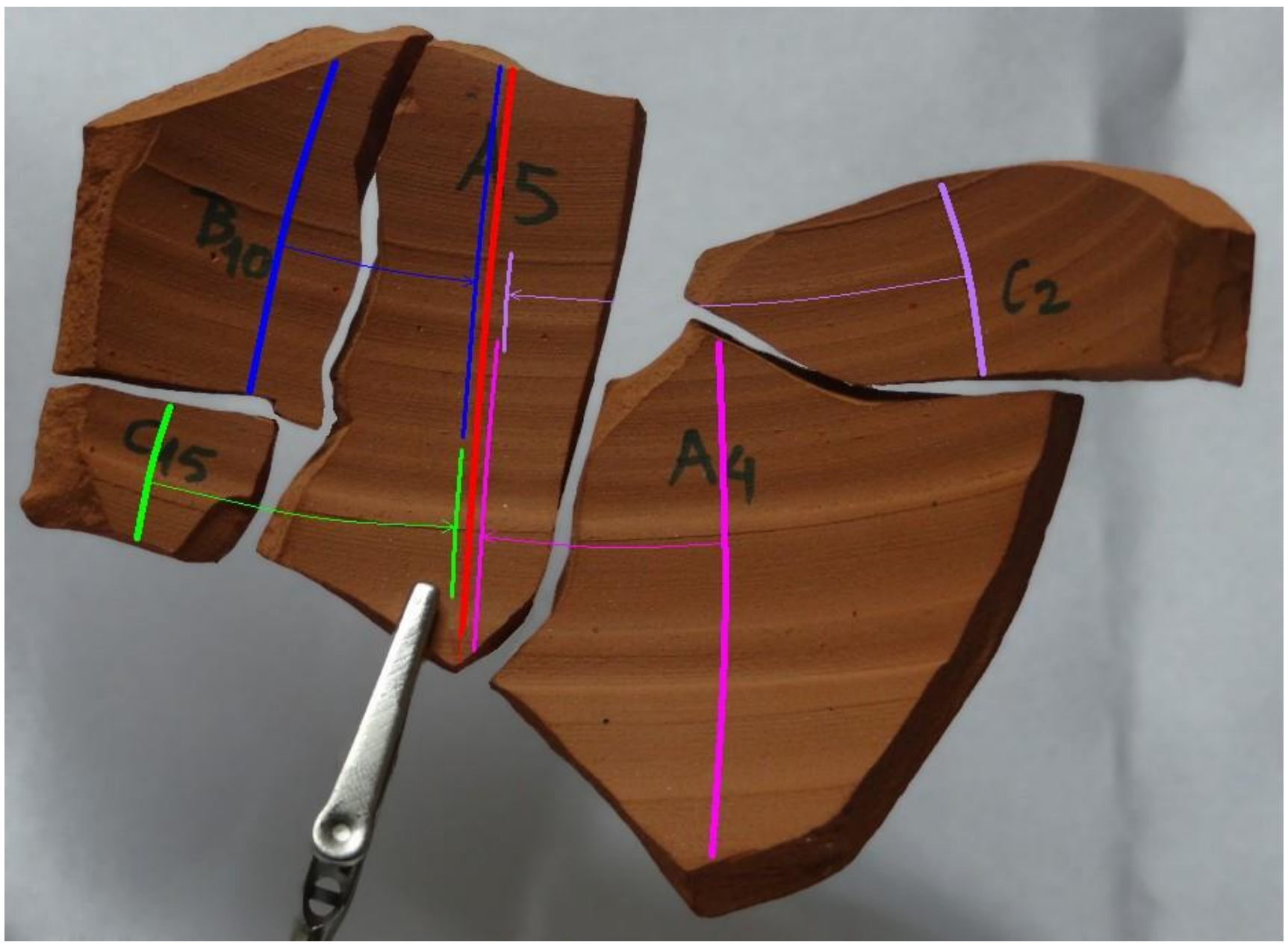

**Fig. 10.** The back surface of the example sherds after the reassembly and the matching of their selected profile planes.

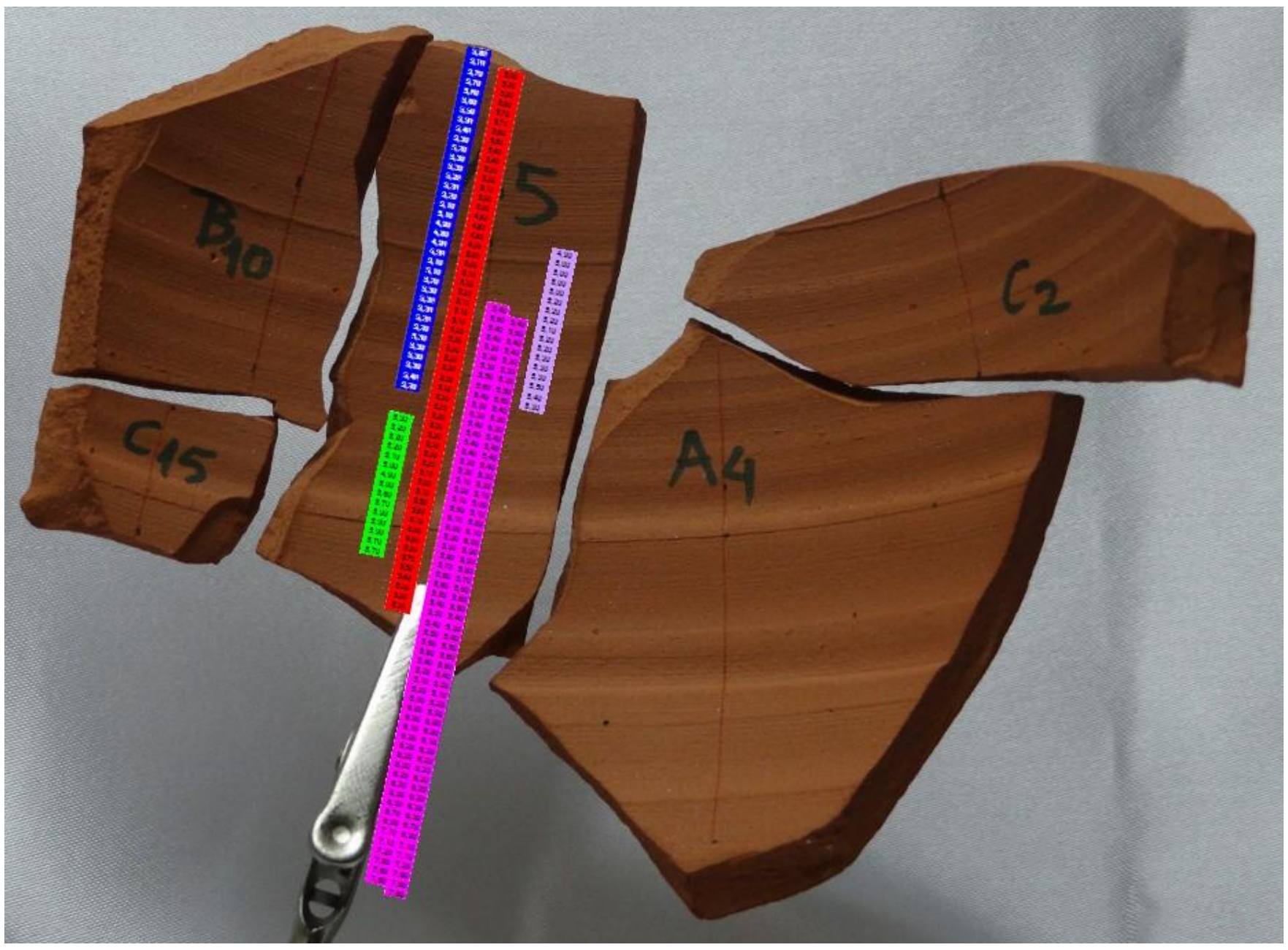

**Fig. 11.** The back surface of the example sherds after the reassembly and the respective Thickness Profiles (accurate thickness measurements).

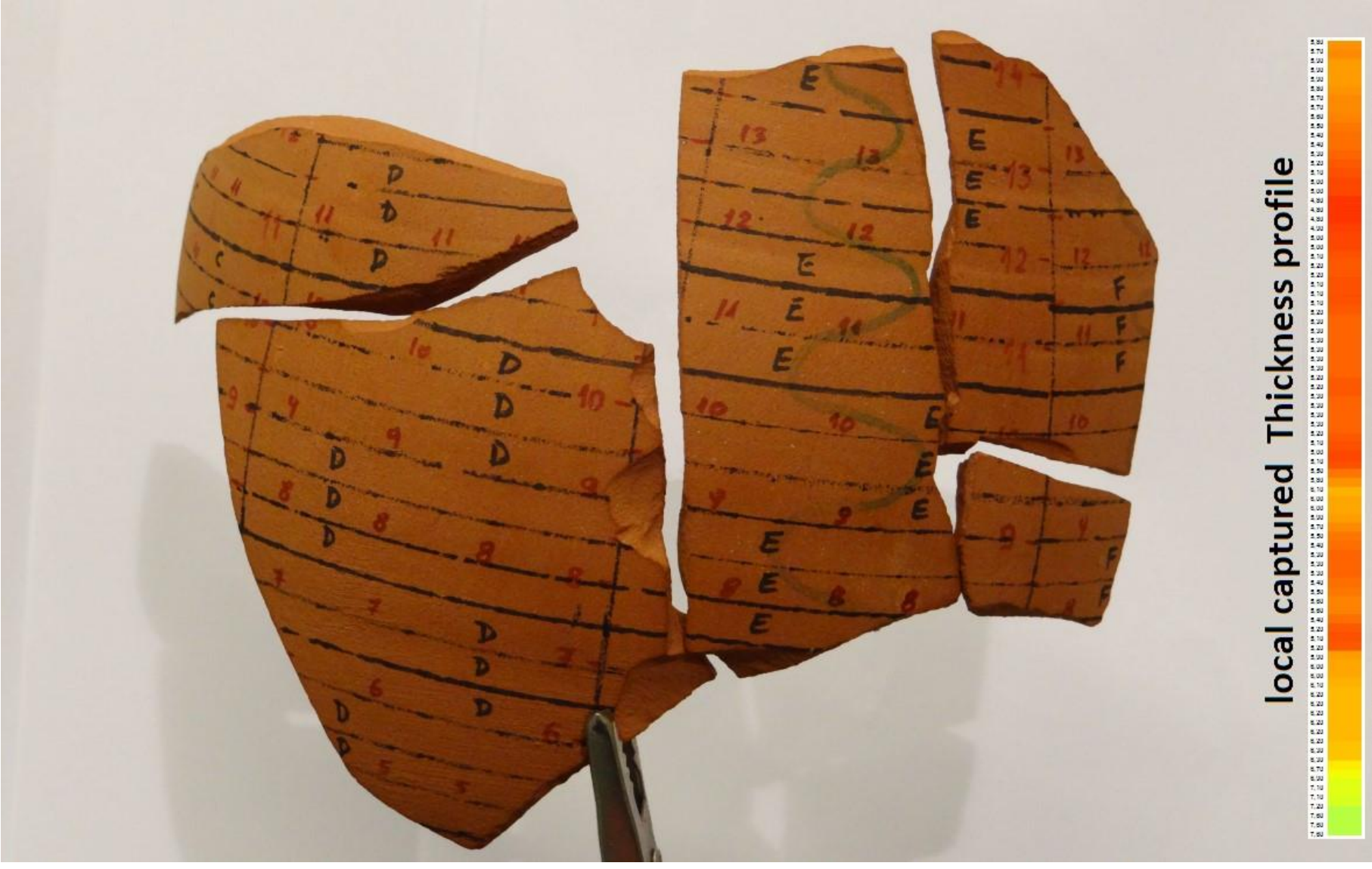

**Fig. 12.** The final reassembled external surface of the pottery along with a thickness scale.